\documentclass[final]{elsarticleARXIVE}

\usepackage{graphicx,epsfig,rotating,color}

\usepackage{amssymb}
\usepackage{amsmath}

\begin{document}

\begin{frontmatter}



\title{Do firms share the same functional form of their growth rate distribution? A new statistical test}


\author[a]{Jos\'e. T. Lunardi}
\ead{jttlunardi@uepg.br}

\author[b]{Salvatore Miccich\`e}
\ead{micciche@unipa.it}

\author[b,c,d]{Fabrizio Lillo}
\ead{lillo@unipa.it}

\author[b]{Rosario N. Mantegna\corref{*}}
\cortext[*]{Corresponding author: {\texttt{mantegna@unipa.it}} (Rosario N. Mantegna)}

\author[e]{Mauro Gallegati}
\ead{mauro.gallegati@univpm.it}

\address[a]{Departamento de Matem\'atica e Estat\'{\i}stica,
Universidade Estadual de Ponta Grossa\\Avenida General Carlos
Cavalcanti, 4748. 84032-900 Ponta Grossa, PR, Brazil}

\address[b]{Dipartimento di Fisica, Universit\`a di Palermo\\ Viale Delle Scienze, Ed 18. 90128 Palermo, Italy}

\address[c]{Santa Fe Institute, 1399 Hyde Park Road, 87501, Santa Fe, NM, USA}

\address[d]{Scuola Normale Superiore di Pisa, Piazza dei Cavalieri 7, 56126 Pisa, Italy}

\address[e]{Dipartimento di Economia, Universit\`a Politecnica Delle Marche\\ Piazza Martelli, 8. 60121 Ancona, Italy}

\begin{abstract}
{We introduce a new statistical test of the hypothesis that a balanced panel of firms have the same growth rate distribution or, more generally, that they share the same functional form of growth rate distribution. We applied the test to European Union and US publicly quoted manufacturing firms data, considering functional forms belonging to the Subbotin family of distributions. While our hypotheses are rejected for the vast majority of sets at the sector level, we cannot rejected them at the subsector level, indicating that homogenous panels of firms could be described by a common functional form of growth rate distribution.}
\end{abstract}


\end{frontmatter}


\section{Introduction}
\label{intro}

Since R. Gibrat  proposed a stochastic model to describe the growth of a firm \cite{Gibrat1931} an important branch of the literature has been concerned with the empirical testing of its consequences (see \cite{Sutton1997} for a review and also \cite{Santarelli} for updated references). Gibrat's model deals with the growth of a single firm and it is based upon two assumptions, namely (i) the \emph{Law of Proportionate Effect} (also known as \emph{Gibrat's law}), stating that the proportionate growth of a firm in a given period is a random variable independent of the initial firm size and (ii) the assumption of statistical independence of successive growths. The main consequences of the model are that, after a long period (typically one year), the logarithmic growth rates are normally distributed and independent of the initial firm's  size. In the literature the size of the firm is measured with different proxies such as the total annual sales, the total annual revenue/turnover, the number of employees, etc.

Recently, the empirical investigations on the validity of Gibrat's model or of alternative growth models are receiving an increasing and renewed interest motivated by the electronic availability of extensive data sets containing a large number of firms which could in principle allow to scrutinize among alternative models with high statistical accuracy \cite{Stanley1996,Bottazzi2001,Bottazzi2003,Bottazzi2003b,Lotti2003,Fujiwara2004,Bottazzi2006,Bottazzi2006b,Stanley2006,Stanley2008,japan}.
However even in such data sets it is hard to empirically test the single firm models because the available data sets typically contain a large number of firms which are sampled over a small number of time periods each. Moreover for balanced panels, the longer is the time period considered, the smaller is the number of firms in the panel\footnote{Balanced panels contains only firms for which there are data in the whole period under study. As a consequence, firms that enter or exit the market in that period are not considered.}.

The traditional approach to circumvent the above difficulty of short time series is to assume that the growth time series of each individual firm  is a specific realization of the same stochastic process which governs the growth dynamics of all the firms. In other words, such an approach assumes that each firm is statistically identical to the \emph{model firm} (MF) described by that stochastic process. Conversely, the statistical properties of the model firm at a given instant of time can be inferred from the statistical properties of the panel of all the firms at the same time. Moreover, if the MF stochastic process is stationary, the statistical properties of each firm and therefore of the pooled sample of all the firms is time independent. By assuming the MF hypothesis, earlier empirical investigations have in general corroborated Gibrat's model, whereas many recent studies carried over large data sets claim that it must be rejected \cite{Sutton1997, Santarelli}. For instance, several recent works find a non Gaussian, ``tent shaped" distributions for the aggregate of growth rates  \cite{Stanley1996,Bottazzi2001,Bottazzi2006,Bottazzi2006b}, and also evidences in some data sets for a dependence of the growth rate distributions on the firms initial size (see, for example, \cite{Stanley1996}-\cite{Stanley2008} and references therein). These tests of the Gibrat's model are based on the MF hypothesis and it is therefore important to investigate whether this hypothesis is valid.

As we have discussed above, the shortness of the firm growth time series makes difficult to test the MF hypothesis. In this article we introduce a method to test the MF hypothesis, which is powerful even when the time series are short, provided the number of firms in the panel is large, as explained in section \ref{powertest}. We also consider the more general hypothesis that there is some level of heterogeneity among individual firms. More specifically we  also test the hypothesis that all firms in a balanced panel have the same specific functional form of the growth rate distribution, although the parameters that characterize the distribution may be different from firm to firm.  We apply our method to balanced panels of (i) firms homogeneous at the manufacturing sector level and (ii) homogeneous at the lower level of  manufacturing subsectors. We consider different time periods length ranging from 8 to 18 years. We find that the statistical validation of the null hypotheses strongly depends on the homogeneity (in time and in the sectorial composition) of the panel.

The article is organized as follows. In the next section we present the rationale of our approach in testing a functional form common to all firms of a panel. In Section \ref{applications} we apply this approach to the study of balanced panels containing publicly quoted manufacturing firms from Compustat and Amadeus databases. We consider both sector and subsector homogeneous panels of firms. We test the model firm hypothesis and the hypothesis of a specific common functional form of the growth rate distribution by considering distributions belonging to the Subbotin family. In the last section we present our conclusions.

\section{A procedure to test whether the same functional form of the growth rate distribution is common to all firms of a panel}  \label{approach}

In this work we consider only balanced panels containing $N$ firms whose size $S_i^j$ ($i=1,2,\cdots,N$) is recorded for $T+1$ instants of time $j$ ($j=0,1,2,\cdots,T$). The logarithmic growth rate of the $i$-th firm at time $j$ ($j>0$) is defined as
$r_i^j=\ln \frac{S_i^j}{S_i^{j-1}}$. Our basic assumptions are:
\begin{itemize}
\item[B1:] single firms grow independently among themselves;
\item[B2:] the time series of the growth rate of each firm consists of statistically independent records;
\item[B3:] the time series of growth rate of each firm is stationary.
\item[B4:] let $p_i(r_i)$ be the time independent probability density function (pdf) of the growth rate $r_i$ for the $i$-th firm. The idiosyncratic parameters of firm $i$ are its mean $\mu_i$ and standard deviation $\sigma_i$ of its growth rate distribution. The pdf of the standardized variable $Z_i=\frac{r_i-\mu_i}{\sigma_i}$ is $p_i^{\mathrm{std}}(z_i)$. We assume that such distribution is the same for all firms and it will be denoted $p^{\mathrm{std}}$.
\end{itemize}
Under assumptions B1-B4 the pdf of the growth rate $r$ obtained by pooling together all firms is given by
\begin{equation}
\label{dprob}
P(r)=\frac{1}{N}\sum_{i=1}^{N} p_i(r).
\end{equation}
$P(r)$ is defined as the pdf of the growth rate $r$ by a random selection of a firm in the pooled sample.
Now, suppose one knows \emph{a priori} the actual idiosyncratic parameters $\mu_i$ and $\sigma_i$ of each firm and standardizes each firm growth rate $r_i$ by using these parameters. The pdf for the aggregate standardized growth rates $Z_i$, analogous to that given in (\ref{dprob}), will be given by $P_Z(z)=\frac{1}{N}\sum_{i=1}^{N} p_i^{\mathrm{std}}(z)=p^{\mathrm{std}}(z)$, i.e., it will be identical to the probability distribution assumed to be common to all the single firms. Such a standardization procedure has the effect of removing the idiosyncrasies among single firms. As a special case, if the single firm growth rates $r_i$ are all identically distributed according to a pdf $p_r(r)$ (this case corresponds to the MF hypothesis) then not only $P_Z(z)=p_r^{\mathrm{std}}(z)$, but also $P(r)=p_r(r)$, as it can be straightforwardly seen from Eq. (\ref{dprob}).

The above considerations may suggest that under assumptions B1-B4 it is sufficient to observe the aggregate distribution of all the standardized empirical growth rates to infer the common shape $p^{\mathrm{std}}$ of the single firm growth distributions. Unfortunately, in practical situations things are not so simple. The main difficulty is that one does not know \emph{a priori} the actual idiosyncratic parameters $\mu_i$ and $\sigma_i$. Usually these parameters must be estimated from the empirical time series of each individual firm. As already mentioned, available data sets typically contain a small number of time records for each firm and consequently only very imprecise estimates for these parameters are available. Such noisy estimates of the parameters may cause the aggregate distribution of standardized growth rates to deviate significantly from the theoretically expected $p^{\mathrm{std}}(z)$.

\begin{center}
\begin{figure}
\includegraphics[scale=0.22]{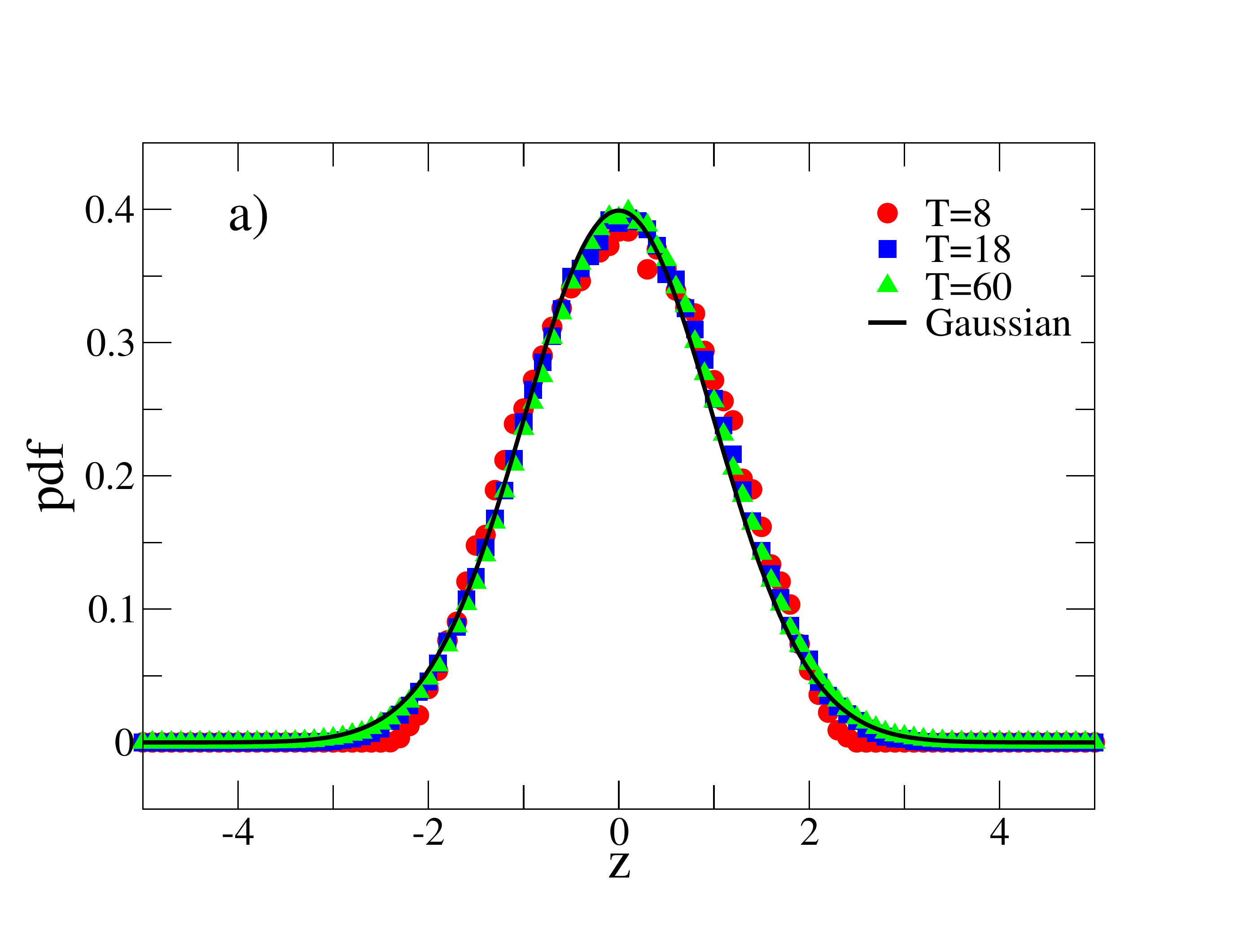}
\includegraphics[scale=0.22]{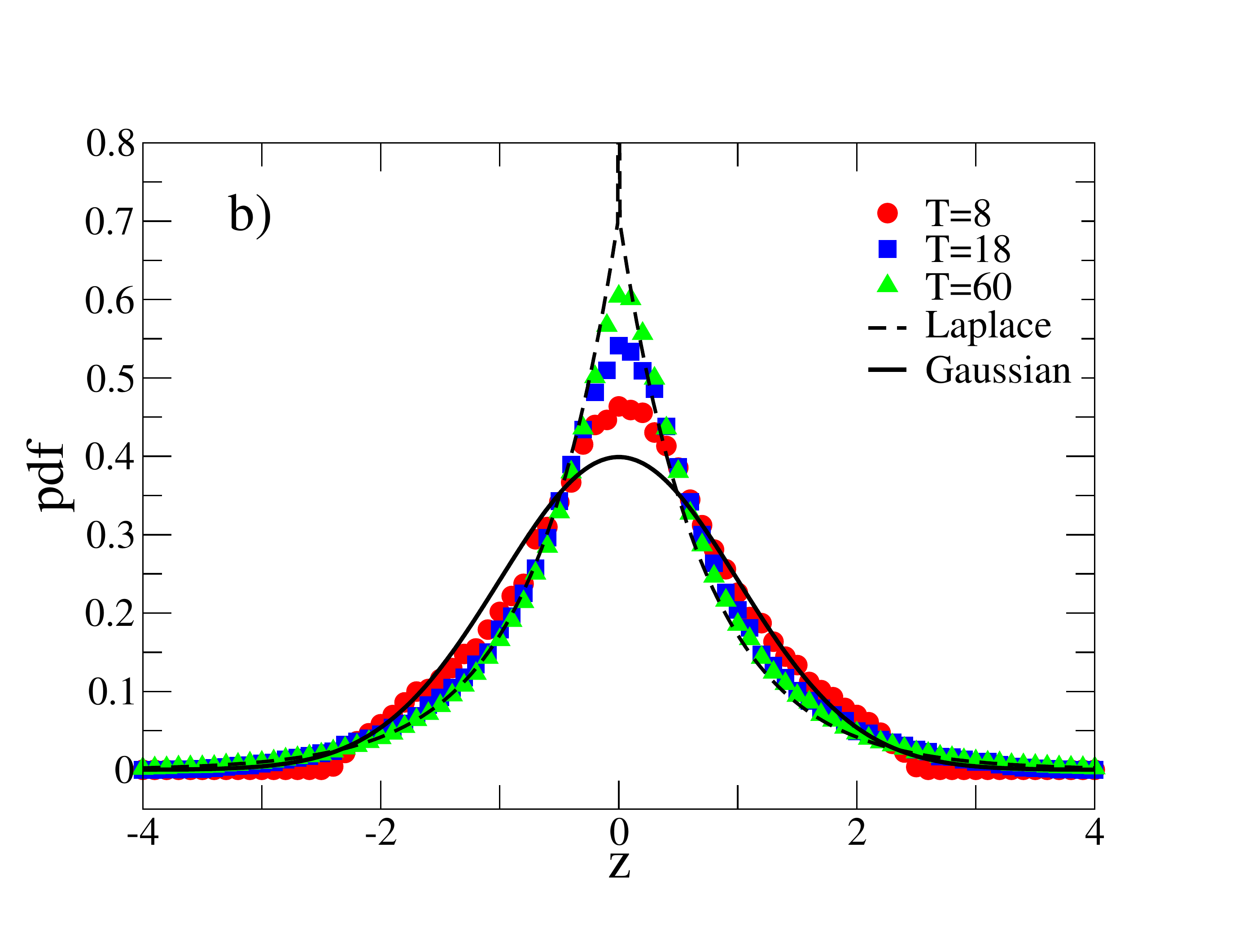}
\caption{\scriptsize{ (Color)  Histograms for the aggregate of all standardized artificial data for several time series lengths. Data are generated according to a Gaussian (panel a) and Laplace (panel b) distribution. In each figure we plotted  for comparison the corresponding distribution $p^{\mathrm{std}}(z)$ which would be the expected one if the mean and standard deviation estimates were sufficiently accurate. In panel b} we plotted also the standard Gaussian distribution, for the sake of comparisons. \label{fig1}}
\end{figure}
\end{center}

In order to illustrate this point we perform some numerical simulations.  Denote the empirical time series of growth rates of the $i$-th firm  by $\{r_i^j\}_{j=1}^{T}$, and the time series estimates of the parameters $\mu_i$ and $\sigma_i$ by $\hat{\mu}_i$ and $\hat{\sigma}_i$, respectively.  By standardizing each observed single firm time series by using these sample estimates we obtain the corresponding time series of the empirically standardized growth rates $\{z_i^j\}_{j=1}^{T}$ for the $i$-th single firm, where $z_i^j=\frac{r_i^j-\hat{\mu}_i}{\hat{\sigma}_i}$. From now on we shall refer to this standardization with sample moments as a {\it z-transformation} and the corresponding standardized growth rate as {\it z-growth rate}.
We artificially generated  $N=10,000$ independent time series of length $T$, each one generated from $T$ independent outcomes drawn from a given distribution. Each time series was standardized by using the usual sample estimates of the mean and standard deviation. We then plot the histogram for the aggregate of all standardized artificial data. We consider different time series length ranging from $T=8$ to $T=60$. These values are typically observed in empirical investigations of real firm data.
As it is illustrated in Figure \ref{fig1}, the distribution of the aggregate z-transformed outcomes may be quite different from the distribution we used to generate the data. Such a deviation is more significant for short time series length $T$. Whereas for time series generated from a Gaussian distribution (see Fig \ref{fig1}a) the z-transformation does not change  appreciably the functional form, for non Gaussian variables the use of sample moments to perform the standardization can dramatically change the distribution.
For example, in Fig \ref{fig1}b we observe that for time series of length $T=8$ generated from a Laplace distribution, the distribution of the aggregate standardized variables
looks more similar to a Gaussian than to a Laplace distribution. Thus, from a visual inspection of this plot, one could be lead to incorrectly conclude that data are generated from a Gaussian distribution. Customary statistical tests corroborate our conclusion that the standardization procedure with sample moments on short time series lead to aggregate standardized variables whose distribution is artificially close to a Gaussian.
This example illustrates the need for an accurate test for discriminating among alternative hypotheses for the common functional form of the growth rate distribution.

Since for short time series the distribution of z-growth rate obtained by z-transforming with sample moments may be very different from the underlying common functional form, one should compare the empirical z-growth distribution with a distribution which takes into account the deformations caused by the noisy nature of parameters estimators. Our idea is to find this expected distribution through a Monte Carlo procedure.

Suppose we have an empirical set of $N$ firm growth rates recorded for $T$ time steps. We want to test the hypothesis that a specific functional form $p^{\mathrm{std}}(z)$ is common to all firms and that firms are different only because the mean and the standard deviation of the growth rate is idiosyncratic for each firm. The steps of our test are the following:
\begin{itemize}
 \item[1.] if one wants to test a specific functional form for the single firms (a Gaussian, a Laplacian or any other specific functional form), then generate a large number $M$ ($M\gg N$) of surrogate time series of length $T$, $\{y_k^j\}_{j=1}^{T}$, ($k=1,...,M$), independently drawn from the hypothesized single firm distribution. Without loss of generality we can take such a distribution to be $p^{\mathrm{std}}(z)$ for all the artificial time series (this statement will become clear after step 2). On the other hand, if one wants to test the MF hypothesis, for which the single firm distribution is not known, we first built an $N \times T$ data matrix ${\cal{D}}={d_{ij}}$ where each entry $d_{ij}$ is the growth rate  $r_i^j$ of the $i$-th firm at the $j$-th time period, with $i=1, \cdots, N$ and $j=1, \cdots, T$. We then build $M$ new surrogate firm time series by randomly choosing (with replacements) $T$ values from matrix ${\cal{D}}$;

 \item[2.] z-transform each of these $M$ artificial time series by using its corresponding estimates ${{\hat{\mu}}_y}{_{_{k}}}$ and ${{\hat{\sigma}}_y}{_{_{k}}}$ for the mean and standard deviation, respectively\footnote{A sample of outcomes $\{y^j\}_{j=1}^{T}$ from a pdf $p_Y$ with parameters $\mu$ and $\sigma$ can be obtained by first sampling a set $\{x^j\}_{j=1}^{T}$ from the standard version of $p_Y$ and, subsequently, by performing the transformation $y^j=\mu+\sigma x^j$. The z-transformed sample corresponding to the $y$ outcomes will then be $\{z_y^j\}_{j=1}^{T}$, with $z_y^j=\frac{y^j-{\hat{\mu}}_y}{{\hat{\sigma}}_y}$, where ${\hat{\mu}}_y=\mu+\sigma {\hat{\mu}}_x$ and ${\hat{\sigma}}_y=\sigma {\hat{\sigma}}_x$, and ${\hat{\mu}}_x$ and ${\hat{\sigma}}_x$ are the time series estimates for the mean and standard deviation of the outcomes $\{x^j\}_{j=1}^{T}$, respectively. Therefore, $z_y^j=\frac{\left(\mu+\sigma x^j\right)-\left(\mu+\sigma {\hat{\mu}}_x \right)}{\sigma {\hat{\sigma}}_x}=\frac{x^j-{\hat{\mu}}_x}{{\hat{\sigma}}_x}=z_x^j$. In other words, after step 2 the artificially generated z-growth rates would be insensitive with respect to any arbitrary set of ``artificial idiosyncratic parameters" $\mu_k$ and $\sigma_k$ ($k=1,\cdots,M$) one could associate to the set of $M$ artificial time series.};

 \item[3.] construct the distribution of all the above artificial z-growth rates pooled together. This is the distribution of  aggregate of z-growth rates expected under the null hypothesis;

 \item[4.] apply several goodness of fit tests to compare the expected distribution obtained in step 3 with that obtained from the empirical z-growth rates.
\end{itemize}

In what concerns the last step we shall follow reference \cite{edf} and use goodness of fit tests based on the empirical cumulative distribution function (EDF tests). As EDF tests require the expected cumulative distribution obtained under the null hypothesis to be continuous, in this work we did linear interpolations between each successive pair of discrete points of the discrete distribution obtained in step 3 in order to make it suitable for the use in such tests.

\subsection{Analysis of the power of the test for a class of distributions} \label{powertest} 

Before we apply our approach to analyze empirical data we present some studies concerning the power of our test. For the last step of the procedure introduced above we shall consider, as suggested in \cite{edf}, the statistics $A^2$ (Anderson-Darling), $W^2$ (Cramer-Von Mises), and $U^2$ (Watson), besides $D$ (Kolmogorov). We will reject the null hypothesis at the level of significance $\alpha$ if it is rejected by at least one of these four statistics at that significance level. This global test (considering all the four EDF statistics combined) tends to increase the probability of a Type I  Error (the probability to reject the null when it is valid) beyond the value $\alpha$. On the other hand, with this criterion we expect to increase the power of the global test in comparison with each single test applied separately.

In order to study the power of the global test we used a Monte Carlo procedure to generate a thousand of $N\times T$ panels, each of them containing $N$ independent time series of length $T$ drawn from a specified distribution $H_a$ (the ``alternative" hypothesis) and after \emph{z}-transformed by using the time series estimates. We built panels with $N$ ranging from $5$ to $350$, and for $T=9$ and $T=18$. As for the alternative hypothesis we considered some members of the Subbotin family of distributions, which  includes the Laplacian and the Gaussian distributions as particular cases and whose pdf has the following general form
\begin{equation}
p(r)=\frac{1}{2\gamma\beta^{\frac{1}{\beta}}\Gamma\left(\frac{1}{\beta}+1\right)}
\exp\left(-\frac{1}{\beta}\left|\frac{r-\mu}{\gamma}\right|^\beta\right)\, .\label{subbotin}
\end{equation}
The parameter $\beta$ characterizes the shape of the distribution, the parameter $\mu$ is the mean and, for each $\beta$, parameter $\gamma$ is proportional to the standard deviation. In fact the standard deviation is given by $\sigma=\gamma\,\beta^{\frac{1}{\beta}}\sqrt{\frac{\Gamma\left(\frac{3}{\beta}\right)}{\Gamma\left(\frac{1}{\beta}\right)}}$. The smaller is the value of the shape parameter $\beta$, the more leptokurtic is the distribution. For the alternative distributions we choose the Subbotin members given by $\beta=1/2$, $3/4$, $1$ (Laplace), $3/2$, and $2$ (Gaussian).

Through a Monte Carlo procedure we then calculate the critical values of each of the four considered statistics at the level of significance $\alpha$ for a set of null hypotheses identical to that chosen for the alternatives $H_a$ (for the details about the Monte Carlo procedure to calculate critical values for EDF statistics see \cite{fagiolo}).

Finally, we applied the global test in each of the $1,000$ panels to calculate the rate of rejection of the null $H_0$ for each case (each case corresponds to a given $H_a$ and a given panel dimension $N\times T$). When $H_0\neq H_a$ this rate corresponds to the power of the global test, whereas when $H_0=H_a$ such a rate corresponds to the effective probability of a Type I  Error.

\begin{center}
\begin{figure}
\includegraphics[width=12.8cm,height=9.5cm]{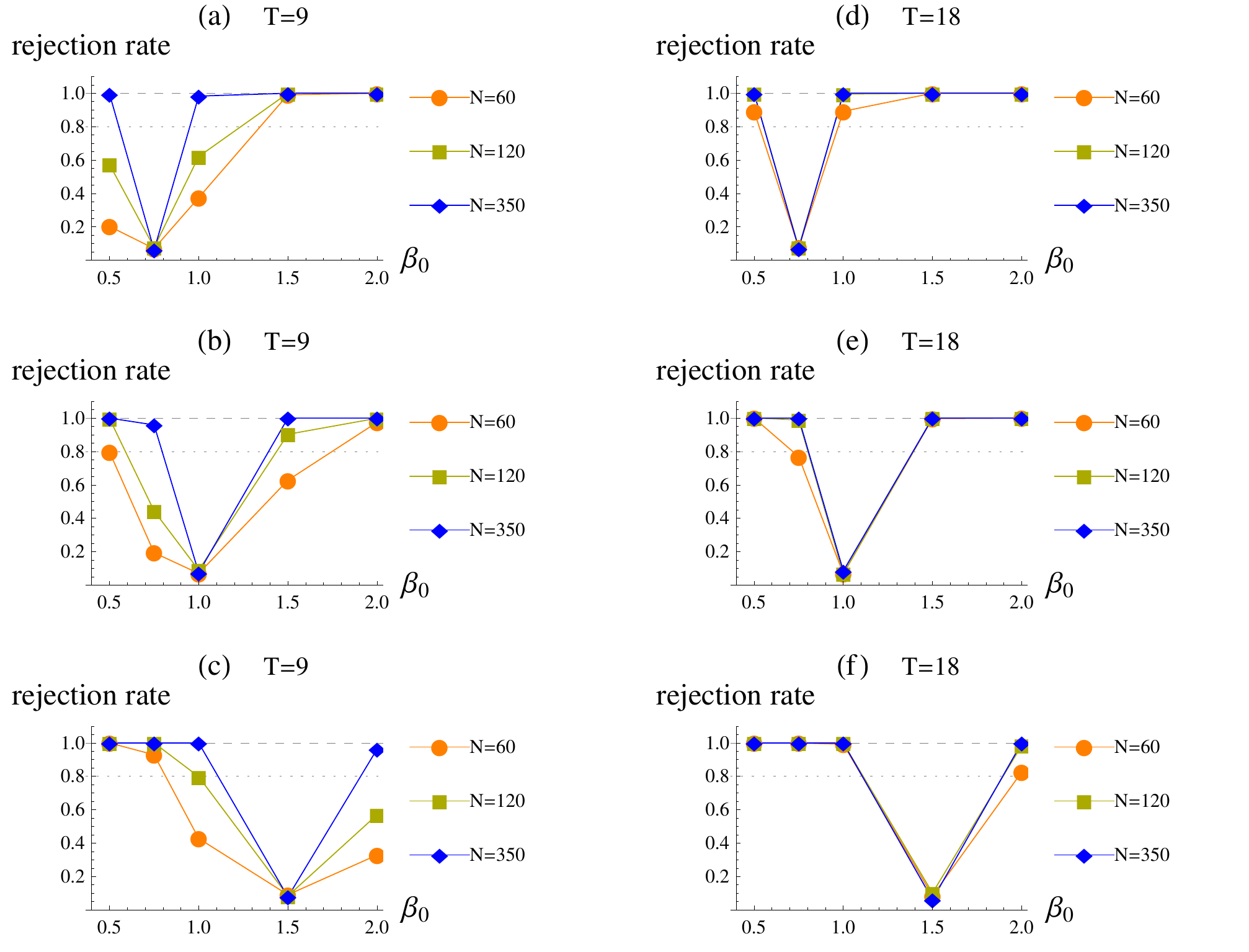}
\caption{\scriptsize{ (Color) Analysis of the power of  the global test for a significance level $\alpha=0.05$. The horizontal axis corresponds to the values of the shape parameter $\beta_0$, associated with the null $H_0$. The alternative distribution $H_a$, from which the data were actually drawn, is a Subbotin with $\beta=3/4$ [plots (a) and (d)], $\beta=1$ [plots (b) and (e)], and $\beta=3/2$ [plots (c) and (f)]. Horizontal dashed and dotted lines at rejection rates of respectively $100\%$ and $80\%$ are shown for reference. In each plot the value of the minimum gives the effective probability of a Type I Error in the global test (which is in general slightly greater than $\alpha$).} \label{newfig2}}
\end{figure}
\end{center}

Figure \ref{newfig2} shows the results for some of the cases studied, but it illustrates the general behavior for all the cases we considered\footnote{An enlarged set of results is available upon request}. In that Figure we can observe that when $T$ and $N$ are both relatively small the test shows a low power to reject nulls which are close to the actual distribution $H_a$. For instance, plot 2(a) shows that when $H_a$ is a Subbotin with $\beta=3/4$ there is a high probability that the nulls corresponding to $\beta=1/2$ and $\beta=1$ will be accepted if $N$ is relatively small. For fixed $N$ and $T$ the power increases as $H_0$ deviates from $H_a$. On the other hand, if $N$ \emph{or} $T$ is relatively large the power is very high in general. For a high value of $T$ and a not so small value of $N$ ($N>60$, for instance), the test have a very high probability to reject a null which deviates even slightly from $H_a$. For instance, plots for $T=18$ [(b), (d), and (f)] show that the valleys are narrower in comparison with those in the plots at the first column, in such a way that even the nulls closer to the actual $H_a$ have a high probability to be rejected.

\section{Statistical tests of empirical data}                \label{applications}

In this section we illustrate our approach by applying it to six data sets. Data were obtained from two major databases: (i) the {\emph{Compustat} database, and (ii) \emph{Amadeus Top 250,000} database.

The Compustat database \cite{Compustat} is issued by Standard \& Poor's and contains data derived from publicly quoted North American companies. The Amadeus Top 250,000  \cite{Amadeus} is issued by Bureau van Dijk and contains data about European firms\footnote{To be included a company should satisfy at least one of the following size inclusion criteria. For the UK, Germany, France, Italy, Spain, Ukraine and Russian Federation: operating revenue equal to at least 15 million EUR, total assets equal to at least 30 million EUR, number of employees equal to at least 200. For all other countries:  operating revenue equal to at least 10 million EUR, total assets equal to at least 20 million EUR, number of employees equal to at least 150. Note that companies with ratios Turnover per employee or Total Assets per employee below 1,000 EUR are excluded from the TOP 250,000}. The Amadeus Top 250,000 database also gives the information about whether or not a firm is publicly quoted. In order to be consistent with respect to the Compustat database, in our investigations we considered only those firms from the Amadeus Top 250,000 database that are publicly quoted. For both databases we considered only manufacturing firms according to the NAICS classification. Specifically, the manufacturing sector is characterized by 2-digit NAICS codes ranging from 31 to 33 \cite{naics}.

Specifically, the six data sets we investigated are
\begin{itemize}
\item[{D1:}] all the European Union\footnote{We considered here the 25 countries version for the European Union. Amadeus Top 250,000 has also an European Union version comprising 15 countries.} firms in Amadeus Top 250,000 database for which there are data on  annual revenue/turnover for the whole period from 1996 to 2004 ($N=698$, $T=8$);
\item[{D2:}] all the US firms in Compustat database for which there are data on annual sales for the whole period from 1981 to 1999 ($N=764$, $T=18$);
\item[{D3:}] the same set of firms of D2, but with the time period restricted to 1981-1990 ($N=764$, $T=9$);
\item[{D4:}] the same set of firms of D2, but with the time period restricted to 1990-1999 ($N=764$, $T=9$);
\item[{D5:}] all the US firms in Compustat database for which there are data on annual sales for the whole period from 1990 to 1999 ($N=1,412$, $T=9$);
\item[{D6:}] all firms present in D5 but not in D4, for the same time period common to both data sets ($N=648$, $T=9$).
\end{itemize}
These sets are homogeneous at the manufacturing sector level. In the next sub-section we also apply our test to panels of firms which are homogeneous at the subsector level.

As a result, the proxy of the firms size is given by $(i)$  data expressed in Euro on annual revenue/turnover in set D1 (Amadeus) and $(ii)$ data expressed in dollars on annual sales for all the other sets (Compustat). For each firm, the growth rates thus obtained have then been de-trended for the annual GDP variation of the country in which the firm is located. The data for the GDP variation and the exchange rates to convert the national currencies to Euro (or ECU, before Euro) were obtained from the International Financial Statistics (IFS) of the International Monetary Fund \cite{IMF}.

For each of the above sets of data we tested six null hypotheses. One of them is the MF hypothesis that all firms have a common single firm distribution also characterized by the same parameters for all firms. In the remaining five cases the null hypothesis is that the single firm distribution is given by a member of the Subbotin family of distributions of Eq. $(\ref{subbotin})$ \cite{Bottazzi2003,Fagiolo2008} characterized by a mean and standard deviation that can be different for different firms. We consider the same Subbotin distribution as those used in the previous section for the analysis of test power (i.e. $\beta=1/2,3/4,1,3/2,$ and $2$) \footnote{The investigation can be extended to other classes of distributions, including asymmetric ones.}.  Monte Carlo simulations were used to compute the relevant critical p-values at $1\%$, $5\%$, and $10\%$ significance levels for the considered EDF statistics, according to the procedure illustrated in subsection \ref{powertest}.
\begin{sidewaystable}
\begin{small}
\begin{tabular}{|lc|c|ccccc|}\hline
Data       &    & MF    & $\beta\!=\!1/2$ &$\beta\!=\!3/4$  & $\beta\!=\!1$ (Laplace)&$\beta\!=\!3/2$ &$\beta\!=\!2$ (Gaussian)\\
set        & \emph{S}   &  $\hat{\sigma}_{\mathrm{obs}}\,\,\left(S_{\mathrm{crit}}^{0.01}\right)\,\,\left(S_{\mathrm{crit}}^{0.05}\right)$  & $\hat{\sigma}_{\mathrm{obs}}\,\,\left(S_{\mathrm{crit}}^{0.01}\right)\,\,\left(S_{\mathrm{crit}}^{0.05}\right)$  & $\hat{\sigma}_{\mathrm{obs}}\,\,\left(S_{\mathrm{crit}}^{0.01}\right)\,\,\left(S_{\mathrm{crit}}^{0.05}\right)$  & $\hat{\sigma}_{\mathrm{obs}}\,\,\left(S_{\mathrm{crit}}^{0.01}\right)\,\,\left(S_{\mathrm{crit}}^{0.05}\right)$  & $\hat{\sigma}_{\mathrm{obs}}\,\,\left(S_{\mathrm{crit}}^{0.01}\right)\,\,\left(S_{\mathrm{crit}}^{0.05}\right)$ & $\hat{\sigma}_{\mathrm{obs}}\,\,\left(S_{\mathrm{crit}}^{0.01}\right)\,\,\left(S_{\mathrm{crit}}^{0.05}\right)$  \\
\hline
&&&&&&&\\
{\textbf{D1}}       &$A^2$ & $   5.99 \, (1.99)\,(1.32)$    & $ 30.5  \, (2.72)\,(1.69)$    & $ 9.80  \, (1.90)\,(1.23)$    & $  3.04 \, (1.49)\,(1.02)$     & $ 2.90 \, (1.24)\,(0.85)$ & $   7.70 \, (1.13)\,(0.81)$   \\
$N\!\!=\!\!698$    &$W^2$ & $   0.85 \, (0.36)\,(0.23)$    & $ 4.74  \, (0.53)\,(0.33)$    & $ 1.57  \, (0.35)\,(0.22)$    & $  0.48 \, (0.26)\,(0.17)$     & $ 0.42  \, (0.20)\,(0.14)$ & $   1.16 \, (0.18)\,(0.12)$  \\
$T\!\!=\!\!8$       &$U^2$ & $   0.85 \, (0.31)\,(0.20)$    & $ 4.65  \, (0.40)\,(0.26)$    & $ 1.51  \, (0.29)\,(0.19)$    & $  0.42 \, (0.22)\,(0.15)$     & $ 0.37  \, (0.18)\,(0.13)$ & $   1.11 \, (0.16)\,(0.12)$    \\
(1996-2004)      &$D$   & $   1.81 \, (1.37)\,(1.14)$    & $ 3.96  \, (1.61)\,(1.31)$    & $ 2.36  \, (1.36)\,(1.12)$    & $  1.51 \, (1.22)\,(1.01)$     & $ 1.36  \, (1.10)\,(0.93)$ & $   2.07 \, (1.05)\,(0.90)$    \\
\hline
&&&&&&&\\
{\textbf{D2}}       &$A^2$ & $   52.0 \, (3.39)\,(2.14)$    & $ 144 \, (4.61)\,(2.81)$    & $ 27.3  \, (2.36)\,(1.57)$    &  $ 2.96 \, (1.63)\,(1.15)$     & $  30.6     \, (1.19)\,(0.84)$ & $  75.2 \, (1.05)\,(0.768)$\\
$N\!\!=\!\!764$    &$W^2$ & $   8.79 \, (0.696)\,(0.430)$    & $  24.7 \, (1.00)\,(0.591)$    & $ 5.05  \, (0.486)\,(0.310)$    &  $ 0.381 \, (0.313)\,(0.213)$     & $ 4.57   \, (0.203)\,(0.142)$& $  12.0 \, (0.178)\,(0.126)$ \\
$T\!\!=\!\!18$      &$U^2$ & $   8.75 \, (0.468)\,(0.303)$    & $ 24.6 \, (0.618)\,(0.389)$    & $ 4.99  \, (0.334)\,(0.230)$    &  $ 0.339 \, (0.243)\,(0.172)$     & $ 4.54   \, (0.180)\,(0.127)$ & $  12.0 \, (0.165)\,(0.118)$ \\
(1981-1999)      &$D$   & $   4.86 \, (1.75)\,(1.42)$    & $ 8.31 \, (2.06)\,(1.66)$    & $ 4.05  \, (1.53)\,(1.27)$    &  $ ^{*}1.21 \, (1.31)\,(1.10)$     & $ 3.38  \, (1.12)\,(0.949)$ & $  5.39 \, (1.05)\,(0.909)$    \\
&&&&&&&\\
{\textbf{D3}} &$A^2$ &  $17.21  \,(2.32) \, (1.45)$  &  $43.78\,  (2.83) \, (1.83) $ &  $12.91\,  (1.94)  \,(1.30) $ &  $2.57 \, (1.47) \, (1.04)$  &  $2.88 \, (1.21) \, (0.84) $ &  $10.76 \, (1.12) \, (0.79)$\\	
$N\!\!=\!\!764$    &$W^2$ &  $2.75 \, (0.44) \, (0.27) $ &  $7.02 \, (0.57) \, (0.36) $ &  $2.18 \, (0.37) \, (0.24) $ &  $0.44 \, (0.26)  \,(0.18) $ & $0.35 \, (0.20) \, (0.14) $  & $1.48\,  (0.18) \, (0.12) $ \\
$T\!\!=\!\!9$      &$U^2$ & $2.69 \, (0.35) \, (0.22)$  & $6.99 \, (0.41) \, (0.27) $ &  $2.15  \,(0.29) \, (0.20)$ & $ 0.42 \, (0.22) \, (0.15) $ & $ 0.33 \, (0.18) \, (0.12)$  & $ 1.47\,  (0.17) \, (0.12)$\\	
(1981-1990)           &$D$   &  $2.97 \,  (1.48) \,  (1.20)$ & $4.55 \,  (1.66) \,  (1.37)$  & 	$2.69  \, (1.39) \,  (1.16)$  & $1.50 \,  (1.22) \,  (1.03)$  &  $1.16 \,  (1.09) \,  (0.93)$  &  $2.17 \,  (1.06) \,  (0.90)$  \\
 &&&&&&&\\
 {\textbf{D4}}       &$A^2$ & $   6.81 \, (2.31)\,(1.47)$    & $ 26.3  \, (2.83)\,(1.83)$    & $ 6.47  \, (1.94)\,(1.30)$    &  $ 3.91 \, (1.47)\,(1.04)$     & $ 14.2  \, (1.21)\,(0.836)$ & $  28.3 \, (1.12)\,(0.795)$   \\
$N\!\!=\!\!764$    &$W^2$ & $   1.17 \, (0.439)\,(0.273)$    & $ 4.46  \, (0.574)\,(0.361)$    & $ 1.20  \, (0.366)\,(0.241)$    &  $ 0.600 \, (0.261)\,(0.178)$     & $ 1.94 \, (0.198)\,(0.136)$ & $  4.02 \, (0.176)\,(0.124)$ \\
$T\!\!=\!\!9$      &$U^2$ & $   1.12 \, (0.343)\,(0.219)$    & $ 4.28 \, (0.411)\,(0.273)$    & $ 1.05  \, (0.289)\,(0.198)$    &  $ 0.472 \, (0.221)\,(0.155)$     & $ 1.84  \, (0.179)\,(0.126)$ & $  3.93 \, (0.167)\,(0.118)$ \\
(1990-1999)           &$D$   & $   2.01 \, (1.46)\,(1.21)$    & $ 3.73 \, (1.66)\,(1.37)$    & $ 2.41  \, (1.39)\,(1.15)$    &  $ 1.67 \, (1.21)\,(1.03)$     & $ 2.55  \, (1.10)\,(0.930)$ & $  3.63 \, (1.06)\,(0.901)$    \\
&&&&&&&\\
{\textbf{D5}}       &$A^2$ & $   31.7 \, (2.69)\,(1.69)$    & $  52.4 \, (2.84)\,(1.87)$    & $ 9.92  \, (1.99)\,(1.29)$    & $  ^{**}0.998 \, (1.61)\,(1.06)$     & $  14.6     \, (1.23)\,(0.851)$ & $  37.5 \, (1.16)\,(0.805)$    \\
$N\!\!=\!\!1412$    &$W^2$ & $   5.07 \, (0.535)\,(0.329)$    & $  8.56  \, (0.568)\,(0.368)$    & $ 1.77  \, (0.376)\,(0.235)$    & $  ^{**}0.114 \, (0.286)\,(0.183)$     & $  1.95  \, (0.203)\,(0.136)$ & $  5.38 \, (0.185)\,(0.126)$    \\
$T\!\!=\!\!9$      &$U^2$ & $   4.94 \, (0.399)\,(0.256)$    & $  8.56 \, (0.418)\,(0.276)$    & $ 1.77   \, (0.298)\,(0.191)$    & $  ^{**}0.111 \, (0.238)\,(0.160)$     & $  1.95  \, (0.184)\,(0.125)$ & $  5.38 \, (0.175)\,(0.120)$    \\
 (1990-1999)          &$D$   & $   3.78 \, (1.62)\,(1.31)$    & $ 4.83  \, (1.66)\,(1.37)$    & $ 2.21  \, (1.39)\,(1.16)$    & $  ^{**}0.921 \, (1.24)\,(1.04)$     & $  2.57  \, (1.10)\,(0.931)$ & $  4.04 \, (1.08)\,(0.902)$    \\
&&&&&&&\\
{\textbf{D6}}       &$A^2$ & $   16.8 \, (2.64)\,(1.65)$    & $ 32.4 \, (2.98)\,(1.89)$    & $ 9.66   \, (1.99)\,(1.32)$    &  $ 3.26 \, (1.54)\,(1.04)$     & $ 6.58 \, (1.24)\,(0.843)$ & $  15.4 \, (1.12)\,(0.806)$    \\
$N\!\!=\!\!648$    &$W^2$ & $   2.52 \, (0.518)\,(0.317)$    & $ 5.33 \, (0.595)\,( 0.373)$    & $ 1.72      \, (0.370)\,(0.241)$    &  $ 0.603 \, (0.269)\,(0.181)$     & $ 1.03  \, (0.197)\,(0.136)$ & $  2.33 \, (0.179)\,(0.126)$    \\
$T\!\!=\!\!9$      &$U^2$ & $   2.51 \, (0.388)\,(0.247)$    & $ 5.13 \, (0.436)\,(0.282)$    & $ 1.58  \, (0.295)\,(0.197)$    &  $ 0.502 \, (0.231)\,(0.156)$     & $ 0.952  \, (0.175)\,(0.127)$& $  2.28 \, (0.168)\,(0.120)$    \\
(1990-1999)           &$D$   & $   2.84 \, (1.56)\,(1.29)$    & $ 3.69 \, (1.69)\,(1.38)$    & $ 2.49  \, (1.39)\,(1.15)$    &  $ 1.87 \, (1.22)\,(1.04)$     & $ 1.78  \, (1.09)\,(0.926)$ & $  2.54 \, (1.05)\,(0.903)$    \\
\hline
  \end{tabular}
\caption{\scriptsize{Observed values $\hat{\sigma}_{\mathrm{obs}}$ of the four EDF statistics for each version of the null hypothesis considered. In parenthesis we indicated $\hat{\sigma}_{\mathrm{crit}}^{0.01}$ and $\hat{\sigma}_{\mathrm{crit}}^{0.05}$, which give respectively the $1\%$ and $5\%$ significance level critical values for each statistic. The null hypothesis is rejected at a significance level $w$ if $\hat{\sigma}_{obs}>S_{\mathrm{crit}}^{w}$}.   One asterisk indicates the cases where the test is passed with a $p-$value between 1\% and 5\%, and two asterisks indicate the cases where the test is passed with a $p-$value at the significance level greater than 5\%.}
\label{tab1}
\end{small}
\end{sidewaystable}

\subsection{Results on panels homogenous at the sector level}           \label{results}

We summarize all the results in Table \ref{tab1}. We first note that for all the data sets, all the four EDF statistics reject the MF hypothesis with a $p$-value $<1\%$ for all data sets (see the first column of results in Table \ref{tab1}). This means that, if B1-B4 are valid assumptions, then the individual firms are heterogeneous with respect to the growth rate mean and standard deviation.

The five remaining versions for the null hypothesis tested whether those distributions share a common functional form. For all the data sets and all the considered statistics, one must reject a common functional form of the distribution corresponding to  $\beta=1/2$, $\beta=3/4$, $\beta=3/2$, and $\beta=2$ (Gaussian), with a $p$-value $<1\%$. For data sets D1, D3, D4, and D6 the Laplace distribution ($\beta=1$) is also rejected by all the statistics with a $p$-value $<1\%$.

On the other hand, for data set D5 all the tests show that one cannot reject the null hypothesis of Laplace distribution at a significance level of $5\%$. 
For data set D2 the Laplace null hypothesis cannot be rejected according to the Kolmogorov test with a $p$-value between $1\%$ and $5\%$, and it is rejected by the remaining statistics ($p$-value $<1\%$).

Data sets D4 and D6 are disjoint and their union is the D5 set of firms. The selection of the D4 and D6 sets is based on their past business activity. In fact firms in D4 were active in the whole period 1981-1999 and therefore have a business activity documented well before the starting data of recording in the panel (1990), whereas those in D6 were active in a shorter time sub-period before 1990. Our results suggest that the common Laplace functional form observed in D5 might depend on the specific way the panel is selected, since the same statistical conclusions are not shared by both its subsets D4 and D6 which are discriminating about the business history of the firms. Figure \ref{fig2}a shows the histogram corresponding to data set D5, as well as the expected pdfs  for the six null hypotheses considered here. A visual inspection corroborates the results of Table \ref{tab1}, in the sense that the Laplace pdf ($\beta=1$) fits the empirical distribution better than the other pdfs. On the other hand, Figure  \ref{fig2}b illustrates the heterogeneity between the two subsets D4 and D6 of set D5. It is clear that set D4 and D6 are asymmetric with respect to zero and their asymmetry is opposite. When one consider set D5 these opposite asymmetries cancel out. It is worth noting that the main source of discrepancy of D4 and D6 sets comes from the central part of the distribution whereas the tails are well described by the Laplace distribution.
\begin{center}
\begin{figure}
\includegraphics[scale=0.22]{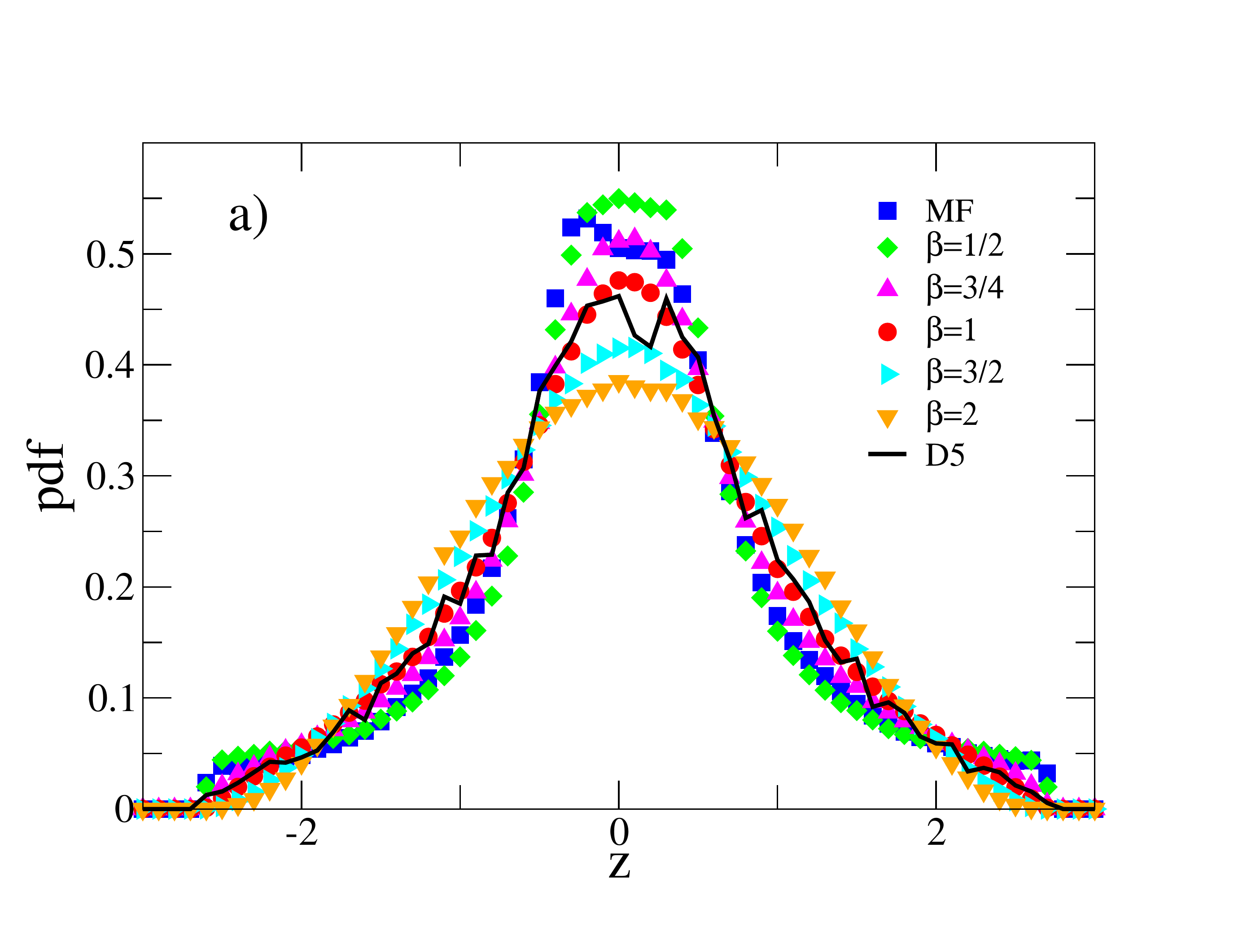}
\includegraphics[scale=0.22]{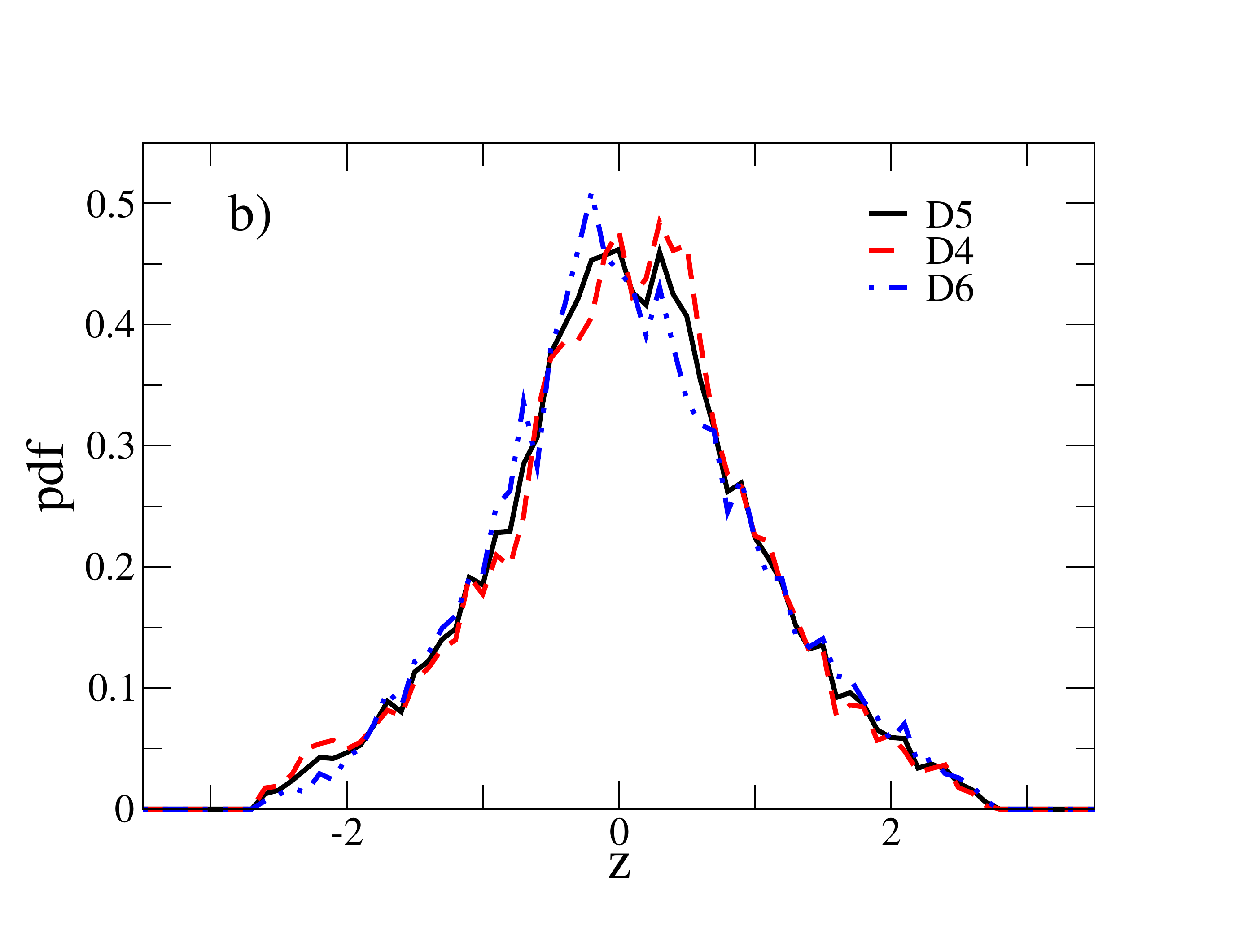}
\caption{\scriptsize{(Color) \emph{a}) Comparison between the pdf of the aggregate of empirical z-growth rates for set D5 and the pdf for the six null hypotheses considered. \emph{b}) Pdf of the aggregate of empirical z-growth rates for data set D5 and its subsets D4 and D6.}\label{fig2}}
\end{figure}
\end{center}

\subsection{Results on panels homogenous at the subsector level}             \label{sec:subsectors}

We now consider panels of firms which are homogeneous at the subsector level. The role of subsector homogeneity of panel of firms in the growth rate distribution has been considered, for example, in \cite{Bottazzi2006,Bottazzi2006b}. The NAICS classification for both databases contains a three digits classification of firms in terms of subsectors. Since most subsectors in our set of firms are small, here we focus on three large manufacturing three digit subsectors for each of the data sets considered above. Specifically for the European Union (Amadeus) set we consider subsectors Chemical Manufacturing (code 325) Computer and Electronic Product Manufacturing (code 334), and Food Manufacturing (code 311). For the US (Compustat) database we consider the same first two subsectors and Machinery Manufacturing (code 333).

We perform the same statistical tests as in the sector case of the previous section, namely we use the EDF statistics to test the model firm hypothesis and the hypothesis that the growth rate of different firms is described by the same functional form belonging to the Subbotin class of Eq. (\ref{subbotin}) with $\beta=1/2, 3/4, 1, 3/2$, and $2$. The result of these statistical tests is summarized in Table \ref{subsectors}. For a given set ($D1,...,D6$) and for a given sector/subsector, we report in boldface those cases for which all the four EDF statistics accept the hypothesis  with a $p$-value larger than 10\%. We report also those less stringent cases for which all the four EDF statistics accept the hypothesis  with a $p$-value larger than 5\%.

As already mentioned above, at the manufacturing sector level only data set D5 is able to satisfy the criteria for the inclusion in the table. For three digit subsectors the results are pretty different. In fact, we find that for most panels there is one or more hypotheses that are accepted according to our strict test. For example, for subsector 325 of set $D3$ both the Laplace ($\beta=1$) and the Subbotin with $\beta=3/2$ hypothesis are accepted with an high $p$-value. The results of subsection \ref{powertest} indicate that this might be due to the low power of our test to discriminate among nulls which are somewhat close to the actual empirical distribution (this is a typical case of small T and relatively small N). Besides that, we recall that in this case the power of the test increases as the null deviates from the actual distribution. Taking into consideration such features of the test, we could guess that the actual distribution may be well approximated by some Subbotin with $\beta$ close to $1$. Moreover, overall our results show that firms homogenous with respect to the subsector of activity are typically well described by a common functional form.
The MF hypothesis is also accepted in several cases sometimes together with a specific functional form (see, for example, panel D6 subsector 333), indicating the possibility that the considered panel is described by a specific functional form with the same idiosyncratic parameters for all firms.
Note also that short time series pass the test more often than long time series. In fact, for set $D2$, describing US firms in the 18 year time period, only subsector 325 passes the test, and for the other two subsectors no null hypothesis is accepted. When we split this set in two 10 year subperiods (set $D3$ and set $D4$)  we accept one or more hypotheses for all the subsectors. Finally, the table shows that the Laplace distribution is the null hypothesis which is accepted more often.

In conclusion our statistical tests give indications that those panels of firms which are homogeneous with respect to the subsector of activity and within a short period of time (roughly a decade) can be often described by some member of the Subbotin family as a common functional form for all the individual firms.  On the other hand, application of our tests to panels which are homogeneous only at the sector level of activity, or to panels encompassing a large time span (roughly two decades), often rejected the nulls considered here.
It is worth mentioning that the results already shown in Fig. \ref{newfig2} indicate that such behaviour might be partly expected due to a diminished power of our test when T is small. On the other hand this effect might also be a genuine one. In fact, it might be related to the fact that (i) different subsectors have sizable idiosyncratic growth components and aggregating subsectors together distorts the form of the distribution; (ii) there are slow time dependent factors which affect the dynamics of the moments of the distribution. Thus even if for short time periods our null hypothesis cannot be rejected, when the investigated period is long the distribution of growth rates is not anymore described by a common functional form; and (iii) for large data sets, or panels  with a large time span, the test has a high power, rejecting nulls which deviate even slightly from the actual distribution. In these cases it could be instructive to consider a more refined set of nulls, perhaps other members of the Subbotin family with $\beta$ interpolating the values considered in this work, or even asymmetric distributions.

\begin{table}
\begin{small}
\begin{center}
\begin{tabular}{||c||c|ccccc||}      \hline \hline
             &  $D1$                                                                   & $D2$                                               & $D3$                                             & $D4$                                             & $D5$                        & $D6$                                           \\
             & {\it{T=8}}                                                             & {\it{T=18}}                                         & {\it{T=9}}                                        & {\it{T=9}}                                      & {\it{T=9}}                    &{\it{T=9}}                                        \\ \hline \hline
Sector & --                                                                           & --                                                       & --                                                    & --                                                    &  Laplace           &--                                                     \\
             & {\it{N=698 }}                                                       & {\it{N=764 }}                                     & {\it{N=764 }}                                 & {\it{N=764 }}                                & {\it{N=1412 }}           &{\it{N=648 }}                                   \\ \hline \hline
325      & Laplace,                                                              &  {\bf Laplace}                                  & {\bf Laplace},                               & $\beta=3/4$,                               &--                                  & --                                                    \\
             &  $\boldsymbol{\beta}${\bf=3/2}                       &                                                           & $\boldsymbol{\beta}${\bf=3/2} & Laplace                                       &                                    &                                                         \\
             & {\it{N=94 }}                                                          & {\it{N=87 }}                                       & {\it{N=87 }}                                   & {\it{N=87 }}                                  & {\it{N=209 }}             &{\it{N=122 }}                                   \\ \hline \hline
334      & {\bf MF}                                                                & --                                                       & Laplace,                                       & {\bf MF}                                        & Laplace                    &{\bf Laplace}                                  \\
            &                                                                               &                                                           &$\boldsymbol{\beta}${\bf=3/2}   &                                                      &                                    &                                                         \\
             & {\it{N=69   }}                                                       & {\it{N=161 }}                                     & {\it{N=161 }}                                 & {\it{N=161 }}                                & {\it{N=364 }}             &{\it{N=203 }}                                   \\ \hline \hline
311     & {\bf MF}, {\bf Laplace},                                      &                                                           &                                                        &                                                      &                                    &                                                         \\
            & $\beta=3/4$,  $\boldsymbol{\beta}${\bf=3/2} &                                                           &                                                        &                                                      &                                    &                                                          \\
             & {\it{N=75 }}                                                         &		                                          &                                                          &                                                     &                                    &                                                       \\ \hline \hline
333     &                                                                               &--                                                        & {\bf MF}                                          &{\bf Laplace},                              & {\bf Laplace},           & MF                                                    \\
            &                                                                               &                                                           &                                                        & $\boldsymbol{\beta}${\bf=3/2}& $\beta=3/2$             &  $\boldsymbol{\beta}${\bf=3/2}   \\
            &                                                                               &                                                           &                                                        &                                                      &                                    &  {\bf Laplace}                                  \\
             &                                                                            & {\it{N=82 }}                                       & {\it{N=82 }}                                    & {\it{N=82 }}                                  & {\it{N=140 }}             &{\it{N=58 }}                                   \\ \hline \hline
\end{tabular}
\caption{Summary of the null hypotheses that are accepted for the different data sets analyzed, at the level of the manufacturing sector and its considered subsectors. We report those hypothesis accepted with a $p$-value larger than 10\% on all four statistics (boldface) and those hypotheses accepted with a $p$-value larger than 5\% on all four statistics (normal). Note that subsector 311 is for European firms only (set $D1$) and subsector 333 is for US firms only (set $D2,...,D6$).}
\label{subsectors}
\end{center}
\end{small}
\end{table}
  	
\section{Conclusions}    \label{conclusions}

We have introduced a method testing hypotheses about the nature of the growth rate stochastic process of single firms belonging to a panel of firms. As a particular case, our method allows to test the widely used model firm hypothesis, which assumes that all firms follow the same stochastic process. The relevance of the test resides in its ability to cope with the common problem in firm growth analysis consisting in having data sets with a large number of firms but  only a small number of time records for each firm.

We have applied our method to empirical data of US and European Union publicly quoted manufacturing firms. Our results indicate that the model firm hypothesis must be rejected for all the considered panels at the sector level of aggregation. Moreover, at this level, and under assumptions B1-B4, our results lead to the rejection of most of our tested hypotheses on a common  functional form of growth rate distribution belonging to the Subbotin family. Among the rejected hypotheses there is the Gaussian distribution, which is the functional form expected if the Gibrat's model were valid at the level of single firms.

In the sector sets considered here, there is also a case of no-rejection (by all the four statistics at the $5\%$ significance level) occurred with set D5 and only for the  null hypothesis of a Laplace distribution, which is often considered in the literature as a valid alternative to the Gaussian one \cite{Stanley1996,Bottazzi2001,Bottazzi2006}.  However, our test rejects the null hypothesis of Laplace distribution for the two subsets of D5, termed D4 and D6.  Specifically, growth rate distribution of firms with longer business activity\footnote{In this paper ``activity" refers to the presence or absence of a firm in the database. A firm that starts the activity in a given year could be either a new firm created in that year or a firm that entered the database in that year because, for example, started to be publicly quoted.} (D4) have more mass for small positive rates, whereas growth rate distribution of firms with a relatively shorter business activity (D6) have more mass for small negative rates. This opposite asymmetry suggests a role of the business activity of a firm which is not captured in the symmetric distributions considered here. Such a conjecture deserves further and deeper investigations.

The investigation of panels of firms belonging to the same subsector indicates that our null hypothesis of a common functional form cannot often be rejected at the 5 \% significance level (in several cases the null cannot be rejected even at the 10 \% significance level). This seems to be more  is much stronger when the investigated time period is not too long (roughly one decade, in our case). Further investigations are needed in order to understand whether such an effect is simply due to a diminished power of our statistical test when $T$ is small, or it is grounded on sound economic motivations.

The Laplace distribution is often accepted in our global statistical test with a significance level of 5 \%, whereas the Gaussian distribution is never accepted at that significance level. Therefore we conclude that for balanced panels of firms homogeneous with respect to the subsector level of activity, and for short time periods, the hypothesis of a common functional form belonging to the Subbotin family of distributions (and sometimes even the model firm hypothesis) typically describes well the growth dynamics of single firms. On the contrary, when one aggregates heterogenous sets of firms (belonging to different subsectors of activity, for instance) or consider large time periods the hypothesis of a common functional form of the growth rate distribution is often rejected.

Finally, we would like to emphasize that our results are relative to balanced panels of publicly quoted firms. One has therefore to consider, for instance, obvious survivorship biases \cite{survivorship} that might prevent a straightforward generalization of our results to a general set of firms.

The results reported in this article confirm that the empirical characterization and the theoretical modeling of the  growth rates distribution of individual firms and of a panel of firms is still an open and challenging subject.

\section{Acknowledgments}
JTL thanks CNPq/Brazil for financial support at the initial stages of this work, as well as the people of OCS/University of Palermo for the kind hospitality. The authors also thank to Prof. Giorgio Fagiolo for very useful discussions about Monte Carlo computation of critical values for EDF tests.


\end{document}